# On the Determination of the Transition to Pure Reptation by Dielectric Spectroscopy

Mengchun Wu[1], Karin J. Bichler[2], Bruno Jakobi[2], Gerald J. Schneider[1,2]

[1] Department of Physics and Astronomy, Louisiana State University, Baton Rouge, LA

[2] Department of Chemistry, Louisiana State University, Baton Rouge, LA

## Abstract

Polymer melts show a characteristic molecular weight dependent relaxation time that can be related to the unentangled and entangled regime. At high molecular weights, influence of contour length fluctuations and constraint release cease and pure reptation prevails. With the broad frequency range dielectric spectroscopy can follow polymer dynamics over a very broad temperature range, with the additional advantage of recording the spectral shape which contains information on reptation. Here we investigate the apparent discrepancy in the molecular weight for the onset of pure reptation from the molecular weight dependence and the spectral shape. We examined the popular derivative method and compared it with a version that includes higher order terms. Higher order terms lead to a more accurate peak shape and position than those determined with the simpler version. This becomes important if experimental spectra contain conductivity and polarization contributions. Higher order terms require the introduction of an interpolating function to analyze experimental spectra, which lets the Havriliak-Negami function appear to be a more robust, yet reliable tool to determine the peak shapes. We reach the conclusion that molecular weight dependence and spectral shape can be both strongly affected by conductivity and polarization contributions. While this leaves uncertainties on the accurate value of the transition molecular weight, the peak shape points to the existence of reptation and contour length fluctuations in polyisoprene with a molecular weight greater than 1000 kg/mol, which would imply a ten times greater threshold molecular weight than expected from previous estimates using the molecular weight dependence.

# Introduction

Fundamental understanding of polymer dynamics is the basic requirement to develop better materials. Despite intense progress in the most recent years, even the apparently simple linear polymer melts are not fully understood. Studies on polymer melts face the challenge of relaxation processes that have extremely broad distribution of relaxation times, $\tau$, and strongly change with temperature, $T$. Dielectric spectroscopy (DS) has proven itself as valuable tool to study relaxation phenomena over a wide temperature and frequency range.[1]

One of the most intriguing challenges is related to reptation which causes a characteristic signature in the molecular weight dependence.[2] From a theoretical point of view, the normal mode peak position which corresponds to the chain end-to-end relaxation time, $\tau$, at least approximately follows $\tau \propto M^2$ for unentangled and $\tau \propto M^3$ for entangled polymer melts[3]. The exponent at low molecular weights is explained by the Rouse model. Reptation leads to $\tau \propto M^3$, whereas experiments on polymers in this high molecular weight region observe a different exponent, $\tau \propto M^4$, associated with contour length fluctuations (CLF) and constraint released (CR).[2,4]

Numerous polyisoprene (PI) melts were studied as a function of the molecular weight by dielectric spectroscopy and a transition to pure reptation at the order of magnitude $\sim 10^2$ kg/mol has been reported.[3] A fit of the molecular weight dependence of the relaxation time indicates a value of $116 \pm 33$ kg/mol (SI), similar to the value of $175 \pm 50$ kg/mol observed from the measurement of the molecular weight dependence of the zero-shear viscosity by rheology.[3,5] However, Glomann et al. analyzed the dielectric spectrum of entangled PI melts and found that the spectral shape of the dielectric loss, $\epsilon''$, can be described by reptation and contour length fluctuations up to very high molecular weights of at least 310 kg/mol.[6] Pilyugina et al. who discussed data of Watanabe drew similar conclusions, experimentally reaching molecular weights up to 308 kg/mol.[7,8] Matsumiya et al. analyzed dielectric spectra to gain more information on constraint release and report data that indicates the existence of CLF at a molecular weight of 1120 kg/mol in case of PI.[8,9] Hence, the spectral shape seems to point to the presence of CLF and CR even at high molecular weights, while the molecular weight dependence suggests the absence of CLF, and CR.[3,5]

This poses the interesting question whether spectral shape and relaxation time are influenced differently by reptation, CLF, and CR or whether additional processes could cause a discrepancy. In fact, dielectric spectroscopy is highly sensitive to conductivity and polarization. Hence, we ask the question whether one of these contributions could cause an apparent contradiction, either in the determination of the relaxation time or the shape. This is highly likely, as the chain relaxation of high molecular weight polymers is slow and often needs to be manually separated from conductivity and electrode polarization effects.

Dielectric spectroscopy measures the complex dielectric permittivity, $\varepsilon^* = \varepsilon' - i\varepsilon''$, with the dielectric storage, $\varepsilon'$, and the dielectric loss, $\varepsilon''$. As both parts are connected by the Kramers-Kronig relation, real and imaginary part contain the same information. Thus, it appears to be very handy to analyze only one portion, e.g., the dielectric loss, because the reciprocal value of the frequency at the peak maximum can be conveniently interpreted as relaxation time, $\tau$.[1] On the other hand, $\varepsilon''$ and conductivity are often superimposed.[1] The archetypical example is the chain end-to-end relaxation of high molecular weight polyisoprene at elevated temperatures.[3] Caused by the high intrinsic conductivity, the peak position of the normal mode can become rather uncertain; thus, relaxation times or power laws may be difficult to obtain.

The analysis of the real part, which is not affected by the conductivity seems to be a viable approach. However, analyzing $\varepsilon'$ is less convenient, as the inflection point corresponds to the relaxation time, which is harder to assess than simply reading a peak position from an apparently well-defined maximum. More importantly, pronounced changes of a power law exponent are less visible in $\varepsilon'$. One potential alternative is to use the negative logarithmic derivative of $\varepsilon'$, which is known to yield a good approximation of the frequency at the peak position of $\varepsilon''$, though the shape may not be well represented, except for certain circumstances. Thus, it seemingly provides at least a very good alternative to analyze the molecular weight dependence of the relaxation time to test whether strong influences of conductivity and polarization could be the origin of the discrepancy in spectrum vs. molecular weight dependence of the relaxation time.

Here, we would like to take advantage of the peak shape for a better understanding of dielectric spectra of well-entangled polymer melts. Hence, we first explore the potential of a better approximation, which at the first glance shows a much better agreement with $\varepsilon''$, but we identify substantial weaknesses in comparison of theory and experiment, which require attention to avoid mistakes in the determination of the peak position and shape of $\varepsilon''$. Surprisingly, the need of interpolating functions for experimental data introduces the empirical Havriliak-Negami function as simple tool to calculate $\varepsilon''$ from $\epsilon'$, independently of the underlying relaxation mechanism and material. It has two important advantages, the simplicity of usage over the mathematical approximation, and compared to the derivative it recovers the peak shape of $\varepsilon''$.

Here, we show the implications by re-analyzing polyisoprene of a few selected molecular weights and comparing these results with literature values. There are four key findings: (i) For low molar masses where the conductivity does not affect $\varepsilon''$ the relaxation times calculated from $\varepsilon'$ and $\varepsilon''$ agree within the experimental accuracy; (ii) For intermediate molecular weights where the conductivity weakly superimposes the peak of the chain end-to-end relaxation, the peak position shows reasonable agreement after careful subtraction of the conductivity from $\varepsilon''$; (iii) At high molecular weights where the conductivity strongly influences the peak, the error of the relaxation time obtained after subtracting the conductivity can be substantial. Surprisingly, the peak shape can be affected to a degree that $\varepsilon'$ is no longer able to recover $\varepsilon''$. However, despite the potential strong influence on the peak shape, the frequency dependence indicates the existence of at least 2 processes, which would be consistent with reptation and CLF being active even at the highest molecular weights. This observation may help to tackle the contradiction of the relaxation time dependence which predicts pure reptation, while the spectral shape has a strong indication of CLF in addition to reptation. (iv) Within the comparison presented below Havriliak-Negami seems to be the most robust and reliable tool, provided that $\varepsilon'$ and $\epsilon''$ are fitted simultaneously with the pre-factor of the polarization calculated with the Kramers-Kronig relation.

# Theoretical

Dielectric spectroscopy measures the complex dielectric permittivity, $\epsilon^* = \epsilon' - i\epsilon''$, with the real, $\epsilon'$, and imaginary part, $\epsilon''$, connected by the Kramers-Kronig relation.[1]

## The Kramers-Kronig Relation

The Kramers-Kronig relation can be written as

$$\epsilon'(\omega) = \epsilon_\infty + \frac{2}{\pi} \int_0^\infty \epsilon''(\zeta) \frac{\zeta}{\zeta^2 - \omega^2} d\zeta \qquad (1)$$

and

$$\epsilon''(\omega) = \frac{1}{\epsilon_0 \omega \rho} + \frac{2}{\pi} \int_0^\infty \epsilon'(\zeta) \frac{\omega}{\omega^2 - \zeta^2} d\zeta \qquad (2)$$

Herein, $\epsilon_0$ is the permittivity of the vacuum, $\omega$ the angular frequency, and $\rho = 1/\sigma_{dc}$ being the specific ohmic resistivity, with $\sigma_{dc}$ as the DC conductivity. As indicated by the conductivity and delineated in the SI, Equations (1) and (2) can also be used in cases in which additional processes contribute only to the imaginary part. Thus, calculating $\epsilon''(\omega)$ from $\epsilon'(\omega)$ could yield a spectral shape unaffected by conductivity, which enters as extra term in Equation (2). As integral equations are sometimes inconvenient a Taylor expansion of the Kramers-Kronig relation may be easier to handle.

## Numerical Approximation of the Kramers-Kronig Relation

For the analysis of discrete experimental spectra a convenient approximation was introduced by Brather[10]

$$\epsilon''(\omega) - \frac{1}{\rho \omega \epsilon_0} = -0.1010 \frac{\partial \epsilon'}{\partial \log \omega} + \sum_{n=1}^\infty \frac{0.4418}{2^n} \left(1 + \frac{1.1238}{4^n}\right) \cdot \left[\epsilon'\left(\frac{\omega}{2^n}\right) - \epsilon'(2^n \omega)\right] \\ \pm 0.0014 \left[\epsilon''(\omega) - \frac{1}{\rho \omega \epsilon_0}\right] \qquad (3)$$

This equation calculates the experimentally measured dielectric loss, $\varepsilon''$, without the conductivity contribution, $\frac{1}{\rho \omega \epsilon_0}$. The right hand-side of Equation (3) uses the logarithmic derivative. Often it is more convenient to use the natural logarithm, connected by $\frac{\partial \epsilon'}{\partial \log \omega} = \ln 10 \cdot \frac{\partial \epsilon'}{\partial \ln \omega} = 2.303 \cdot \frac{\partial \epsilon'}{\partial \ln \omega}$ which leads to

$$\epsilon''(\omega) - \frac{1}{\rho\omega\epsilon_0} = -0.2326 \frac{\partial \epsilon'}{\partial \ln \omega} + \sum_{n=1}^{\infty} \frac{0.4418}{2^n}\left(1 + \frac{1.1238}{4^n}\right) \cdot \left[\epsilon'\left(\frac{\omega}{2^n}\right) - \epsilon'(2^n\omega)\right] \quad (4)$$
$$\pm 0.0014\left[\epsilon''(\omega) - \frac{1}{\rho\omega\epsilon_0}\right]$$

The term $\pm 0.0014\left[\epsilon''(\omega) - \frac{1}{\rho\omega\epsilon_0}\right]$ represents the error estimate of the approximation. The small numerical prefactor indicates an error roughly of 0.1% of $\epsilon''(\omega)$, supposedly much smaller than the uncertainty in experimental values. Hence, it should provide a good approximation for experimental data. For practical reasons only a finite number of summands, $N$, can be calculated. However, as the denominator of the numerical prefactor becomes dominant at high $n$, using the assumption that $\epsilon'$ stays finite may still yield a good approximation. In fact, tests with our data show that the calculations need to be carefully tested to avoid artifacts from a finite sum.

The right-hand side has two terms, the derivative, $\frac{\partial \epsilon'}{\partial \ln \omega}$, and the sum. For sake of convenience, we define

$$\epsilon''_{der} := \frac{\partial \epsilon'}{\partial \ln \omega} \quad (5)$$

and

$$\epsilon''_{sum} := \sum_{n=1}^{\infty} \frac{0.4418}{2^n}\left(1 + \frac{1.1238}{4^n}\right) \cdot \left[\epsilon'\left(\frac{\omega}{2^n}\right) - \epsilon'(2^n\omega)\right] \quad (6)$$

As the approximation was introduced by Brather we abbreviate

$$\epsilon''_{Brather} = -0.2326 \cdot \epsilon''_{der} + \epsilon''_{sum} \quad (7)$$

Two alternative versions of the derivative,

$$\epsilon''_{der,1} = \epsilon'' \propto -\frac{\partial \epsilon'}{\partial \ln \omega} \quad (8)$$

and

$$\left(\epsilon''_{der,2}\right)^2 = (\epsilon'')^2 \propto -\frac{\partial \epsilon'}{\partial \ln \omega} \tag{9}$$

have gained popularity, and both can actually be used to estimate $\epsilon''$.[11-13] As delineated in the SI, Equation (9) is strictly valid for the Debye model,[1, 14]

$$\frac{\epsilon'(\omega) - \epsilon_\infty}{\Delta\epsilon} = \frac{1}{1 + (\omega\tau_D)^2} \tag{10}$$

and

$$\frac{\epsilon''(\omega)}{\Delta\epsilon} = \frac{\omega\tau_D}{1 + (\omega\tau_D)^2} \tag{11}$$

Herein, $\epsilon_\infty = \lim_{\omega \to \infty} \epsilon'(\omega)$, the relaxation strength, $\Delta\epsilon$, and the Debye relaxation time, $\tau_D$, are used.

However, the more general case of experimental data that needs a description by the Havriliak-Negami function[15, 16]

$$\epsilon^*_{HN}(\omega) = \epsilon_\infty + \frac{\Delta\epsilon}{(1 + (i\omega\tau_{HN})^\beta)^\gamma} \tag{12}$$

leads to more complicated considerations. For the sake of calculating the derivative $\frac{\partial \epsilon'}{\partial \ln \omega}$, we separate real and imaginary part. As delineated in the SI information, we obtain[14]

$$\frac{\partial \epsilon'(\omega)}{\partial \ln \omega} = -\frac{\beta\gamma\Delta\epsilon(\omega\tau_{HN})^\beta \cos\left[\frac{\beta\pi}{2} - (1+\gamma)\theta_{HN}\right]}{\left[1 + 2(\omega\tau)^\beta \cos\left(\frac{\beta\pi}{2}\right) + (\omega\tau)^{2\beta}\right]^{\frac{1+\gamma}{2}}} \tag{13}$$

with

$$\theta_{HN}(\omega) = \arctan\left[\frac{\sin\left(\frac{\beta\pi}{2}\right)}{(\omega\tau_{HN})^{-\beta} + \cos\left(\frac{\beta\pi}{2}\right)}\right] \tag{14}$$

The frequency of the peak maximum can be obtained by

$$\omega_{peak} = \frac{1}{\tau_{HN}} \frac{\sin\left[\frac{\beta\pi}{2+2\gamma}\right]^{\frac{1}{\beta}}}{\sin\left[\frac{\beta\gamma\pi}{2+2\gamma}\right]^{\frac{1}{\beta}}} \tag{15}$$

For $\beta = \gamma = 1$ Equation (15) becomes $\omega_{peak} = \frac{1}{\tau_{HN}} = \frac{1}{\tau_D}$, and HN becomes the Debye equation, for which

$$\frac{\partial \epsilon'(\omega)}{\partial \ln \omega} = -2\Delta\epsilon \left[\frac{\omega\tau_D}{1+(\omega\tau_D)^2}\right]^2 = -\frac{2}{\Delta\epsilon}(\epsilon'')^2, \tag{16}$$

which directly verifies Equation (9) and specifies a numerical pre-factor of $2/\Delta\varepsilon$.

Often the Rouse model provides a useful concept to discuss relaxation in polymer melts. For that purpose, we test the compatibility of the Rouse model with the derivatives introduced by Equations (8) and (9).

## Rouse Model

Several authors use Rouse, reptation, Maxwell type, or similar models to describe experimental data[6, 17, 18]. Testing the derivative technique with any existing model that uses a linear combination exceeds the scope of this work. However, the Rouse model acts as simple example that explain the basic idea and potential challenges. Many of them assume the summation of mode contributions. For example, within the framework of the Rouse model analytical calculation of the dielectric permittivity leads to

$$\frac{\epsilon'(\omega) - \epsilon_\infty}{\Delta\epsilon} = \frac{2}{N_R(N_R-1)} \sum_{p=odd}^{N_R-1} \cot^2\left(\frac{p\pi}{2N_R}\right) \frac{1}{1+(\omega\tau_p)^2} \tag{17}$$

and

$$\frac{\epsilon''(\omega)}{\Delta\epsilon} = \frac{2}{N_R(N_R-1)} \sum_{p=odd}^{N_R-1} \cot^2\left(\frac{p\pi}{2N_R}\right) \frac{\omega\tau_p}{1+(\omega\tau_p)^2} \tag{18}$$

In these Equations (17) and (18), we use the relaxation time, $\tau_p$, of the p-th mode. We define $\tau_p$ via the longest relaxation time within the Rouse model, $\tau_p = \tau_R/p^2$. Hence, equations (17) and (18) use a single parameter to describe the spectral shape. We note that in general, the number of Rouse modes, $N_R$, is not identical to the finite number of sum terms used in Equations (3) and (4). The number $N$ in Equations (3) and (4) traces back to a mathematical series expansion, while Equations (17) and (18) use a decomposition of the large-scale polymer dynamics into $N_R$ modes.

The derivative of $\epsilon'$ is:

$$\frac{1}{\Delta\epsilon} \cdot \frac{\partial \epsilon'(\omega)}{\partial \ln \omega} = (-2) \cdot \frac{2}{N_R(N_R - 1)} \sum_{p=odd}^{N_R-1} \cot^2\left(\frac{p\pi}{2N_R}\right) \left(\frac{\omega\tau_p}{1 + (\omega\tau_p)^2}\right)^2 \tag{19}$$

and with

$$\left(\frac{\epsilon''(\omega)}{\Delta\epsilon}\right)^2 = \left(\frac{2}{N_R(N_R - 1)} \sum_{p=odd}^{N_R-1} \cot^2\left(\frac{p\pi}{2N_R}\right) \frac{\omega\tau_p}{1 + (\omega\tau_p)^2}\right)^2 \tag{20}$$

it becomes immediately clear that neither Equation (8) nor Equation (9) can be fulfilled. Hence, in any case when a linear combination is required, the derivative approximation to calculate $\epsilon''$ from $\epsilon'$ appears to be questionable. To draw conclusions for experiments, we compare the different approaches, using calculated data generated with the Debye, Havriliak-Negami, and Rouse like models.

## Model Comparisons

The three columns of Figure 1 illustrate Debye (left), Havriliak-Negami (middle), and the Rouse like (right) models. The advantage of the model calculations is the knowledge of all values, including relaxation strength. The first row shows calculated model spectra, $\epsilon'$ and $\epsilon''$. The figures in the second row compare $\epsilon''$ calculated from $\epsilon'$ using the different approximations, $\epsilon''_{Brather}$ (Equation (4)), $\epsilon''_{der,1} = -0.2326 \, d\epsilon'/d\ln\omega$, $\epsilon''_{der,2} = (-d\epsilon'/d\ln\omega)^{0.5}$, and $\epsilon''_{sum}$, defined by Equation (6). The third row compares model $\epsilon''$ and $\epsilon''_{Brather}$ and indicates the potential numerical error or inaccuracy of the approximations.

In all the exemplified cases of Figure 1, $\epsilon''_{Brather}$, seems to describe the model data represented by $\epsilon''$ very well. All the approximated spectra seem to provide at least a good estimation of the peak frequency of $\epsilon''$. In the experimentally often more complex scenarios, $\epsilon''_{Brather}$, seems to be the appropriate tool, at least from a theoretical point of view. The bottom figures show the errors between original $\epsilon''$ and calculated values, by plotting the difference $\epsilon'' - \epsilon''_{Brather}$, and displaying the error of the numerical estimate as indicated by Equation (4) to be $\pm 0.0014[\epsilon''(\omega) - \frac{1}{\rho\omega\epsilon_0}]$. Here, we neglected the conductivity ($\sigma = 0$). The difference indicates that the error is proportional to $\epsilon''(\omega)$, thus shows the same frequency dependent spectral shape, but the error values are three orders of magnitude smaller, which seems to mirror the accuracy of the numerical approximation. Some oscillations visible in the difference may simply be caused by the accuracy of the numerical calculations but appear to be much smaller than experimental uncertainties and hence are outside the scope of this work.

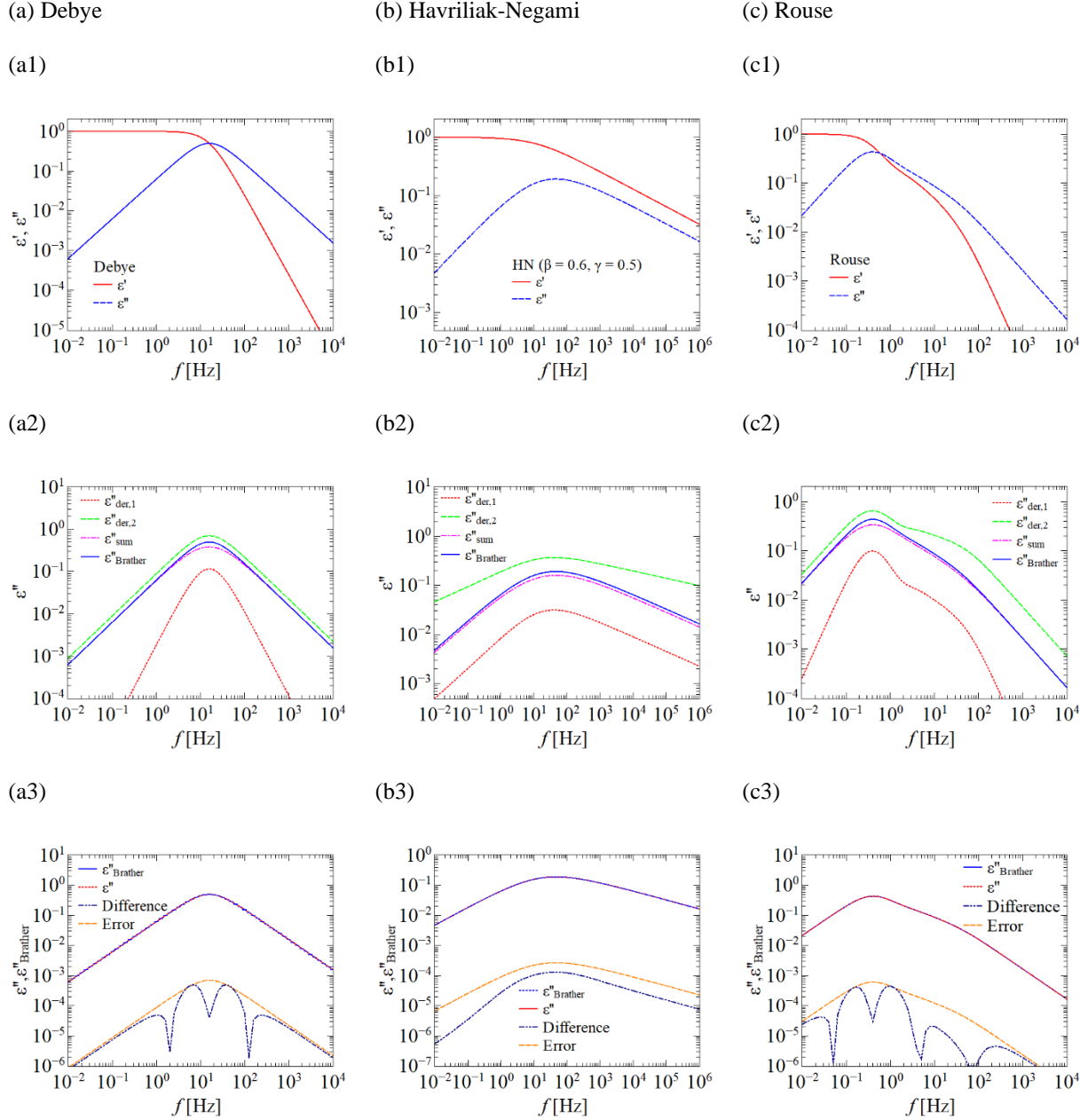

**Figure 1** Columns **(a)** Debye model, **(b)** Havriliak Negami (HN), and **(c)** Rouse model. Line 1 illustrates $\epsilon'$ and $\epsilon''$ for each of the model, line 2 compares the approximations $\epsilon''_{der}$, $\epsilon''_{der}{}^2$, $\epsilon''_{sum}$, $\epsilon''_{Brather}$, and line 3 compares the model $\epsilon''$ with $\epsilon''_{Brather}$ and visualizes the calculated absolute differences between $\epsilon''_{Brather}$ and theoretical $\epsilon''$, and the theoretical error of $\epsilon''_{Brather}$. For the sake of simplicity, the relaxation strength is set to 1 for all models. For the sake of simplicity, $\epsilon_\infty = 0$ and $\Delta \epsilon = 1$ are set for all models. The Debye model used a Debye relaxation time $\tau_D = 0.01$ s. The HN model used ($\beta = 0.6, \gamma = 0.5$) and a HN relaxation time, $\tau_{HN} = 0.01$ s and a Rouse relaxation time, $\tau_R = 0.4$ s and N = 20 modes. (The choice $N = 20$ will be explained in the results and discussion section.)

# Experimental

## Materials

Polymers, 1,4-cis-polyisoprene, with different weight average molecular weights, $M_w$, and polydispersity as specified in Table 1 were purchased from Polymer Standard Service (PSS), Mainz, Germany and used as received. Sample characterization has been done by PSS. If not specified otherwise, we use the abbreviation polyisoprene or PI for the sake of simplicity.

**Table 1** Weight average molecular weight, $M_w$, number average molecular weight, $M_n$, and polydispersity index, PDI, of the samples as determined by Polymer Standard Service (PSS), Mainz, Germany.

| Sample | $M_w$ (g/mol) | $M_n$ (g/mol) | PDI |
|---|---|---|---|
| PI 12000 | 12000 | 11800 | 1.02 |
| PI 231000 | 231000 | 227000 | 1.02 |
| PI 436000 | 436000 | 425000 | 1.03 |

## Experimental Methods and Sample Preparation

The dielectric spectroscopy measurements were performed with a broadband dielectric Alpha analyzer from Novocontrol GmbH with an available frequency range of $f = 10^{-2}$ Hz to $f = 10^6$ Hz. The used temperature range of $T = -100$ °C to $T = 80$ °C, with increments of $\Delta T = 5$ °C, was controlled by a Quatro Cryosystem with an accuracy of 0.1°C by using an evaporated liquid nitrogen stream. For the measurements, the samples were placed between two gold electrodes, with finely cut 0.1 mm thick cross-shaped Teflon pieces as spacer material for ensuring a constant sample thickness. High molecular weight PI is highly viscous and has been annealed at 60 °C overnight under moderate vacuum conditions to avoid cavities between electrodes.

## Data Analysis

As described in the theoretical section, here we are interested in the relaxation time and the spectral shape, which both seem to carry a signature of CLF. The advantage of $\epsilon''$ is the proportionality to $\omega^{-0.25}$ that indicates the existence of CLF. For the sake of simplicity, we use the Havriliak-Negami function to parametrize our data. This parametrization has the primary intention to accurately reflect the data by interpolation of the experimental results, initially with less importance of the parameters.

As our results below illustrate, the dielectric spectra of polyisoprene (PI) with a relatively low molecular weight can be well described with an independent HN function for each the normal mode and the $\alpha$-relaxation. Here we limit ourselves to normal mode and $\alpha$-relaxation. Secondary relaxations are beyond the scope of this work.

At higher molecular weights $M > M_e$, reptation and CLF contribute to the normal mode and lead to an extremely broad spectrum that captures a frequency range of $10^{-2}$ to $10^6$ Hz. Hence a sum of 2 HN functions is needed to accurately describe the spectral shape of the normal mode. The relaxation time of polymers change with the molecular weight, $\tau \propto M^2$ for $M < M_e$ and $\tau \propto M^4$ for $M > M_e$.[3] At least at higher molecular weights, the relaxation time of the $\alpha$-process is molecular weight independent.[1] Hence, we select those temperatures where the $\alpha$-relaxation does not have a significant contribution to the spectral shape, which would involve a 3rd HN term.

Hence our model function reads

$$\epsilon' = \epsilon_\infty + A_1 \omega^{-\alpha} + \text{Re}\left\{\frac{\Delta\epsilon_1}{(1+(i\omega\tau_{HN.1})^{\beta_1})^{\gamma_1}}\right\} + \text{Re}\left\{\frac{\Delta\epsilon_2}{(1+(i\omega\tau_{HN.2})^{\beta_2})^{\gamma_2}}\right\} \quad (21)$$

and

$$\epsilon'' = \frac{A_c}{\omega} + A_2 \omega^{-\alpha} + \text{Im}\left\{\frac{\Delta\epsilon_1}{(1+(i\omega\tau_{HN.1})^{\beta_1})^{\gamma_1}}\right\} + \text{Im}\left\{\frac{\Delta\epsilon_2}{(1+(i\omega\tau_{HN.2})^{\beta_2})^{\gamma_2}}\right\} \quad (22)$$

To be clear here. At low molecular weights HN1 and HN2 describe normal mode and $\alpha$-relaxation, respectively, as both processes are in the frequency window of the dielectric spectrometer. At higher molecular weight the broad normal mode peak and the relaxation time difference between normal mode and $\alpha$-relaxation allow us to concentrate on the normal mode. However, the complex shape requires HN1 and HN2 to describe the complex shape of the normal mode of the high molecular weight polymers.

Equation (22) introduces $A_c$ for the strength of the conductivity. The strength of the polarization is described by the two parameters $A_1$ and $A_2$, in equations (21) and (22), respectively. In the frequency range of our experiments two polarization processes can be distinguished, electrode polarization (EP), and Maxwell-Wagner-Sillars (MWS) polarization.

Here, we study the homopolymer PI for which MWS seems to have a negligible contribution and a single term for EP should suffice.[1] This may not be the case for blends, nanocomposites or microcomposites which may have internal interfaces, as indicated by numerous experiments.[19-23] In case of pure polymer melts interfaces may exist, e.g., free volume can be measured.[24-26] Even within this simplified view, it becomes obvious that the 2 pre-factors of conductivity and polarization in Equation (22) bear some arbitrariness and could simply lead to a wrong estimate of the relative strength of conductivity and polarization contribution.

A potential solution comes from the Kramers-Kronig relation, which can be used to relate $A_1$ and $A_2$. If $\epsilon'$ has a contribution of the form $\epsilon' = A\omega^{-\alpha}$, with $0 < \alpha < 1$, then the corresponding contribution of such power-law in the dielectric loss $\epsilon''$ from Kramers Kronig transformation would be

$$\epsilon''(\omega) - \frac{1}{\epsilon_0 \omega \rho} = \frac{2}{\pi} \int_o^\infty A x^{-\alpha} \frac{\omega}{\omega^2 - x^2} dx = A \tan\left(\frac{\pi\alpha}{2}\right) \omega^{-\alpha} \tag{23}$$

Hence, whenever there is a power law contribution in dielectric spectra, a conversion factor $\tan(\pi\alpha/2)$ can be used. Hence, we set $A_1 = A_p$, and $A_2 = \tan(\pi\alpha/2) A_p$, with the polarization strength $A_p$. Calculation of this integral can be found in the SI. A similar conversion factor has been proposed without derivation to relate the real and imaginary part of the complex conductivity.[27] This equation (23) has two important consequences (i) The Kramers-Kronig relation connects real and imaginary part and reduces the number of 2 free pre-factors to one pre-factor that can be shared between real- and imaginary part. (ii) Whenever a contribution of the form $A_p \omega^{-\alpha}$ is assumed, then the transformation results in $A_p \tan(\pi\alpha/2) \omega^{-\alpha}$. The exponent $\alpha$ is the same for real- and imaginary part. Hence, deviation from a common exponent would indicate a failure of the assumption of $\omega^{-\alpha}$. A potential reason would be a strong conductivity contribution that could hide polarization in $\epsilon''$. As the pre-factor is frequency independent, an indication would be a change of the power law exponent with frequency.

# Results and Discussion

Experimental results on the molecular weight dependence of the relaxation time of polyisoprene (PI) in dielectric spectroscopy suggest a transition to pure reptation[3] and power law fits (SI) indicate it occurs at the molecular weight $116 \pm 33$ kg/mol (SI). The spectral shape points to the existence of contour length fluctuations (CLF) and constraint release (CR) up to 1120 kg/mol.[9] The scientific question is whether this points to the need of improved molecular models or to the need of an improved analysis of dielectric spectra or to both. Here, we use PI of 3 different molecular weights, 12, 231, and 436 kg/mol to test whether the apparent contradiction could be associated with the experimental accuracy.

Dielectric spectroscopy on the low molecular weight PI (12 kg/mol) has the advantage of polarization and conductivity being visible at frequencies well separated from those of the normal mode and $\alpha$-relaxation. In addition, in case of low molecular weights, the normal mode spectrum is less broad, and the relaxation times of both processes are close. Hence, we can find temperatures at which normal mode and $\alpha$-relaxation are not affected by conductivity and polarization within our experimental frequency window. Increasing the molecular weight leads to a broader more complex normal mode peak, slower relaxation times of normal mode, thus, a well-separated $\alpha$-relaxation, but also a stronger influence of conductivity and polarization. In case of PI with a molecular weight of 231 kg/mol, the normal mode peak can be resolved, but at higher molecular weights, 436 kg/mol, the peak is completely hidden at all reasonable temperatures.

## Polyisoprene 12 kg/mol

Figure 2 illustrates the challenges associated with the analysis of the spectra using polyisoprene (PI) with $M_w = 12$ kg/mol, slightly above the entanglement molecular weight $M_e = 3.9$ kg/mol.[28] We chose a temperature of 233 K, because at this temperature both chain end-to-end relaxation (normal mode) and $\alpha$-relaxation fit in the frequency window ($10^{-2}$ Hz $\leq f \leq 10^6$ Hz) of a single experiment. At this temperature and molecular weight, polarization and conductivity do not visibly contribute to $\epsilon'$ and $\epsilon''$. Hence, the spectra are useful to demonstrate first challenges associated with the analysis. Hereafter, if not specified otherwise polarization relates to electrode polarization (EP), as Maxwell-Wagner-Sillars (MWS) polarization seems to be less important for homopolymers like PI.[1]

As illustrated in Figure 2 we used the same models to convert experimental data as exemplified with the model spectra in Figure 1. We calculated $\epsilon''$ from $\epsilon'$ using different approximations, $\epsilon''_{Brather}$ (Equation (4)), $\epsilon''_{der,1} = -0.2326\, d\epsilon'/\,d\ln\omega$, $\epsilon''_{der,2} = (-d\epsilon'/\,d\ln\omega)^{0.5}$, and $\epsilon''_{sum}$, defined by Equation (6). As in the case of the calculated spectra of Figure 1, peak frequencies seem to be well approximated by all of the mentioned approaches, and based on our finds in Figure 1, it is not surprising, that $\epsilon''_{Brather}$ seems to provide the best description. For reasons that will become obvious below, we also used the HN model to describe the data. Here, we analyzed $\epsilon'$ and $\epsilon''$ separately and with simultaneous fits of $\epsilon'$ and $\epsilon''$, sharing the parameters of the real and imaginary parts. For the sake of completeness, we also included the case in which we fit normal mode and $\alpha$-relaxation separately or as the sum of the two contributions. Respective figures can be found in SI. In case of Figure 2, we used either two separate or the sum of two Havriliak-Negami (HN) functions, one for each, the normal mode and the $\alpha$-relaxation to obtain relaxation time, $\tau_{HN}$, and the shape parameters, $\beta$ and $\gamma$ for each process as specified in Tables 2.1, 2.2, and 3. We reach the important conclusion that fitting $\epsilon'$ or $\epsilon''$ separately or simultaneously leads to results that agrees within the experimental accuracy provided the data in the entire experimental frequency range were fitted. The reason for the need of modeling the entire frequency range comes from the close proximity of the normal mode and $\alpha$-relaxation in case of the PI with 12 kg/mol, in which the there is a small overlap that would lead to deviations if not considered, Figure 2 (b). This ceases at higher molecular weights when there $\tau_{NM}/\tau_\alpha$ becomes larger.[1]

**Table 2.1** Parameter for normal mode relaxation of PI 12 kg/mol, by different fitting methods: **(a)** Separate fits of normal mode and $\alpha$-relaxation of only $\epsilon'$. **(b)** Fit of normal mode and $\alpha$-relaxation as sum of two HN functions for $\epsilon'$ only. **(c)** Fit of normal mode and $\alpha$-relaxation as sum of two HN functions for $\epsilon''$ only. **(d)** Fit of normal mode and $\alpha$-relaxation as sum of two HN functions and $\epsilon'$ and $\epsilon''$ simultaneously.

| Approach | $\beta_{NM}$ | $\gamma_{NM}$ | $\beta_{NM} \times \gamma_{NM}$ | $\tau_{NM}$ (s) | $\Delta\epsilon_{NM}$ |
|---|---|---|---|---|---|
| (a) $\epsilon'$ separate | 0.90±0.02 | 0.64± 0.01 | 0.58±0.02 | 0.96±0.03 | 0.062±0.001 |
| (b) $\epsilon'$ whole spectrum | 1.00±0.01 | 0.48±0.01 | 0.48±0.01 | 1.23±0.03 | 0.063±0.001 |
| (c) $\epsilon''$ whole spectrum | 1.00±0.02 | 0.45±0.02 | 0.45±0.02 | 1.28±0.05 | 0.063±0.001 |
| (d) $\epsilon'$ and $\epsilon''$ simultaneous | 1.00±0.02 | 0.44±0.01 | 0.44±0.01 | 1.37±0.03 | 0.064±0.001 |

(a)

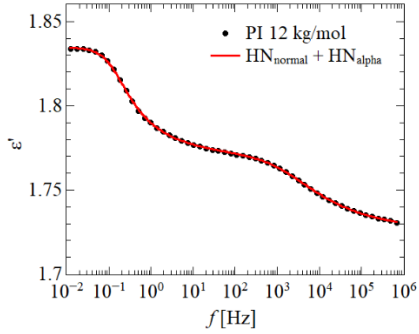

(b)

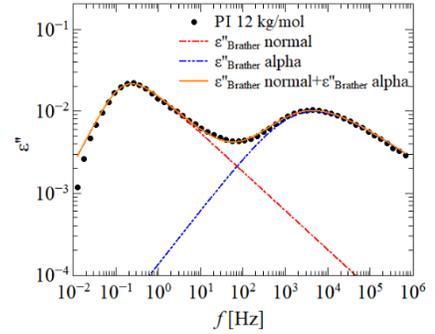

(c)

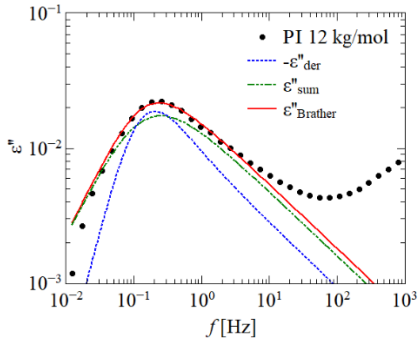

(d)

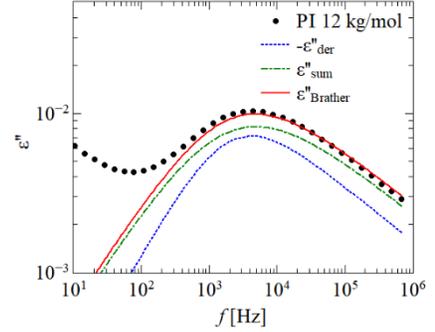

**Figure 2** Dielectric spectrum of polyisoprene (PI) with a molecular weight of 12 kg/mol at a temperature of 233 K. (a) Using Havriliak-Negami (HN) fit of $\epsilon'$. (b) Comparison of HN fit of $\epsilon''$, calculated HN from the parameters of $\epsilon'$, and (c) normal mode and (d) $\alpha$-relaxation. The experimental data is compared with calculated values from $\epsilon''_{der}$, $\epsilon''_{sum}$ and $\epsilon''_{Brather}$.

**Table 2.2** Parameter for $\alpha$-Relaxation of PI 12 kg/mol, by different fitting methods: **(a)** Separate fits of normal mode and $\alpha$-relaxation of only $\epsilon'$. **(b)** Fit of normal mode and $\alpha$-relaxation as sum of two HN functions for $\epsilon'$ only. **(c)** Fit of normal mode and $\alpha$-relaxation as sum of two HN functions for $\epsilon''$ only. **(d)** Fit of normal mode and $\alpha$-relaxation as sum of two HN functions and $\epsilon'$ and $\epsilon''$ simultaneously.

| Approach | $\beta_\alpha$ | $\gamma_\alpha$ | $\beta_\alpha \times \gamma_\alpha$ | $\tau_\alpha$ ($10^{-6}$ s) | $\Delta\epsilon_\alpha$ |
|---|---|---|---|---|---|
| (a) $\epsilon'$ separate | 0.69±0.01 | 0.47±0.01 | 0.32±0.01 | 104±2 | 0.0476±0.0002 |
| (b) $\epsilon'$ whole spectrum | 0.66±0.01 | 0.52±0.02 | 0.34±0.01 | 83±4 | 0.0456±0.0006 |
| (c) $\epsilon''$ whole spectrum | 0.69±0.02 | 0.52±0.03 | 0.36±0.02 | 80±7 | 0.0440±0.0004 |
| (d) $\epsilon'$ and $\epsilon''$ simultaneous | 0.69±0.01 | 0.53±0.01 | 0.37±0.01 | 77±3 | 0.0438±0.0002 |

**Table 3** Relaxation time determined by different approaches for PI 12 kg/mol. Calculated from Table 2 using Equation (15) and from simple manually picking the frequency at which $\epsilon''(\omega) = $ max. The errors of the calculated values have been determined from standard deviations via error propagation and the error of the peak picking results represents the spacing between the experimental frequencies.

| Approach | Normal Mode (Hz) | $\alpha$-relaxation (Hz) |
|---|---|---|
| $\epsilon'$ separate | 0.253±0.008 | 4100±200 |
| $\epsilon'$ whole spectrum | 0.233±0.006 | 4700±300 |
| $\epsilon''$ whole spectrum | 0.236±0.012 | 4700±500 |
| $\epsilon'$ and $\epsilon''$ simultaneous | 0.223±0.008 | 4800±300 |
| Peak picking | 0.25±0.09 | 4400±1500 |

Here, our focus are not the parameters, but the quality of the interpolation of the experimental spectra. As it becomes clear from the figures, the description by $\epsilon''_{Brather}$ appears to be the best approximation within this comparison. However, as it points out, Equation (4) requires the evaluation of $[\epsilon'\left(\frac{\omega}{2^n}\right) - \epsilon'(2^n\omega)]$. Thus, the calculation involves artificial calculation of $\epsilon'$ outside of existing frequency values. Therefore, we turn our attention to a potential influence on the calculated curves. Figure 3 tests the consequence of neglecting all terms $[\epsilon'\left(\frac{\omega}{2^n}\right) - \epsilon'(2^n\omega)]$ outside the frequency window $10^{-2}\text{Hz} \le f \le 10^6\text{Hz}$. It is obvious that a substantial error from such a cut off can be expected. Hence, analyzing the frequency dependence of the spectra by $\epsilon''_{Brather}$ requires careful testing for effects caused by a finite experimental frequency window.

Carefully comparing the experimental results in Figure 2 points to the HN function as simplest, yet most accurate approach that appear to be very robust. As a note, here, the intention is to calculate $\epsilon''$ from $\epsilon'$, for which the empirical character of this function has a lesser meaning.

## Polyisoprene 231 kg/mol

### Fitting $\epsilon'$ only

Knowing that HN provides a convenient, yet accurate means to empirically describe dielectric spectra, we fitted the normal mode of PI with a molecular weight of 231 kg/mol, by the sum of 2 HN functions. As the $\alpha$-

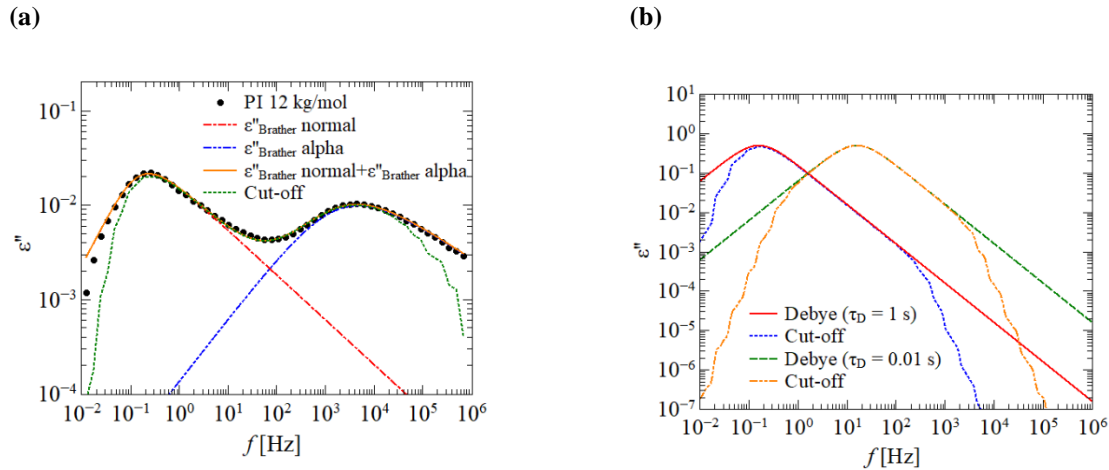

**(a)** **(b)**

**Figure 3** Experimental $\epsilon''$ of PI with a molecular weight of 12 kg/mol compared with calculated values using $\epsilon''_{Brather}$. **(a)** The lines indicate a fit of separate and combined illustration of normal mode and $\alpha$-relaxation. These three curves rely on the assumption that normal mode and $\alpha$-relaxation are known to be outside of the frequency values indicated by the symbols. The dotted line indicates what happens if the frequency values for the calculation of $\epsilon''_{Brather}$ are dropped if outside $10^{-2}$ Hz $\le f \le 10^6$ Hz, i.e., $\epsilon'(\omega) := 0$ **if not** in $10^{-2}$ Hz $\le f \le 10^6$ Hz. **(b)** Simple illustration of cut-off effect on the peak shape using two Debye functions with a different relaxation time $\tau_D = 1$ s and $\tau_D = 0.01$ s.

relaxation is not visible in the experimentally accessible frequency range at the temperature of 343 K, only the normal mode shall be investigated here. A detailed discussion and analysis of the $\alpha$-relaxation can be found in the literature.[1] Fitting our data, it became clear that at least a sum of 2 HN is necessary to describe the normal mode at higher molecular weights. As Figure 4 illustrates, in our first approach, we tried to avoid artificial incorporation of the conductivity by fitting $\epsilon'$ and then calculating $\epsilon''$. The fitting parameters are summarized in Table 4.

The fit of $\epsilon'$ provides satisfactory interpolation of $\epsilon'$ and $\epsilon''$. Deviations at low frequencies of $\epsilon''$ are to be expected as the conductivity contribution intentionally was not included in this consideration. We also left out the polarization contribution to $\epsilon''$ for reasons which will become clear below. It is remarkable that $\epsilon''$ shows power law behaviors $\propto \omega^1$ at the low frequency side of the peak and $\propto \omega^{-0.25}$ at higher frequencies, despite the polarization exponent $\alpha$ (found by fitting $\epsilon'$) goes beyond 1. A purely mathematical conversion of experimental data values does not rely on model assumptions, because it is unrelated to the underlying physics. The result confirms our result based on the HN fit that the peak shape of $\epsilon''$ is apparently overwhelmed by the conductivity contribution. Hereafter, we need to clarify whether it is possible to further quantify conductivity and polarization, which finally would yield a better knowledge of the normal mode spectra, even in the case of strong conductivity and polarization contributions.

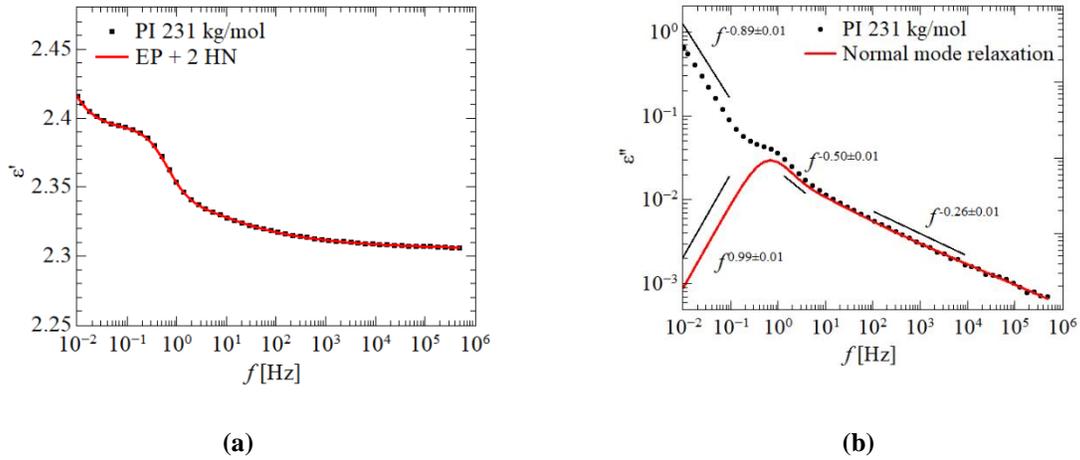

**(a)** **(b)**

**Figure 4** Polyisoprene with a molecular weight of 231 kg/mol at a temperature of 343 K. **(a)** Fitting of experimental $\epsilon'$ by sum of power law and 2 HN functions **(b)** Normal mode relaxation calculated using $\epsilon''_{HN}$ with the parameters from the fit of $\epsilon'_{HN}$ in panel (a). Please note, the polarization has not been included in $\epsilon''$ as the prefactor $A_p \tan\left(\frac{\pi\alpha}{2}\right)$ has been defined for $\alpha < 1$. Hence, we would introduce a free parameter, which is not our intention, cf. text.

**Table 4** Parameters Obtained by fitting the experimental data in Figure 4 with the sum of 2 HN Functions and an electrode polarization contribution based on Equation (21).

|  | $\epsilon_\infty$ | $A_p$ | $\alpha$ |  |
|---|---|---|---|---|
|  | $2.305 \pm 0.001$ | $0.00010 \pm 0.00005$ | $1.18 \pm 0.01$ |  |
| $1^{st}$ HN | $\Delta\epsilon_1$ | $\tau_{HN1}$ (s) | $\beta_1$ | $\gamma_1$ |
|  | $0.060 \pm 0.001$ | $0.258 \pm 0.003$ | $0.999 \pm 0.003$ | $0.91 \pm 0.01$ |
| $2^{nd}$ HN | $\Delta\epsilon_2$ | $\tau_{HN2}$ (s) | $\beta_2$ | $\gamma_2$ |
|  | $0.029 \pm 0.001$ | $0.033 \pm 0.001$ | $0.96 \pm 0.02$ | $0.25 \pm 0.01$ |

To determine the conductivity more accurately, we intend to use the conversion factor $\tan(\pi\alpha/2)$ for the polarization that has been calculated in the experimental data analysis section. As this factor comes from the Kramers-Kronig relation it is accurate and avoids free fit parameters. From this factor it becomes clear that the pre-factor of the polarization in $\epsilon''$ depends on the power law exponent $\alpha$ ($0 < \alpha < 1$). While $\alpha = 0$ would mean a pre-factor equal to zero, the case $\alpha = 1$ would lead to a infinite pre-factor, as $\lim_{\alpha \to 1}(\tan(\frac{\pi\alpha}{2})) = \infty$, and indicate an overwhelmingly strong contribution of the polarization to $\epsilon''$. The value $\alpha = 1.18 \pm 0.01$ found by fitting $\epsilon'$ of Figure 4 a would lead to a potentially wrong overestimate of the polarization contribution to $\epsilon''$, because it is outside the specified range for which the pre-factor was calculated (see section data analysis in the experimental chapter and SI). We can speculate that despite the assumption of the negligible contribution of MWS in homopolymers, a superposition of the two different polarization contributions, EP and MWS, may lead to this observation of a power law exponent > 1. However, our experimental data on PI with 231 kg/mol does not permit to support or disprove this assumption. Hence, we did not include the EP contribution in Figure 4 (b). Instead, we attempt a simultaneous fit using the conversion factor $\tan(\pi\alpha/2)$ and include the conductivity term. This procedure takes advantage of the same information encoded differently into $\epsilon'$ and $\epsilon''$, which avoids overfitting of the imaginary part. As conversions are model independent and reduce the number of fit parameters it can be anticipated that this fit is more robust and reliable.

*Simultaneous Fit of $\epsilon'$ and $\epsilon''$*

Hence, to draw further conclusion, we fitted real and imaginary part simultaneously. As Figure 5 (a) illustrates the fit appears to be less accurate at low frequencies of $\epsilon'$ but surprisingly $\epsilon''$ obtained from either the simultaneous fit or the one calculated from $\epsilon'$ are virtually identical (Figure 5 (c)). Slight differences between the fitting curves from the two approaches occur around the maximum at which there seems to be a strong influence of conductivity and polarization, Figure 5 (b)). In addition, comparing Tables 4 and 5 shows values that are very similar. The most notable difference is the power law exponent of polarization which changes from $\alpha = 1.18 \pm 0.01$ to $\alpha = 0.900 \pm 0.002$. We note here that the indicated error represents the standard deviation provided by the fit, but not the model error, which may be substantially larger as indicated by the deviations in Figure 5 (a). Following Kramers-Kronig relation, $\alpha = 0.9$ leads to a substantial polarization pre-factor (Figure 5 (d)) for $\epsilon''$. As the mathematical calculation avoids free parameters for the polarization it seems to be a more reliable approach to determine the pre-factor of the conductivity term. Surprisingly, the polarization appears to be stronger than the conductivity contribution, Figure 5 (b).

(a) 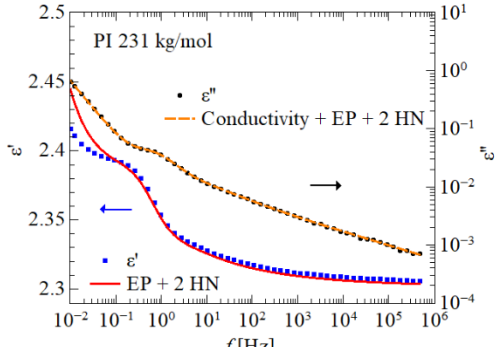

(b) 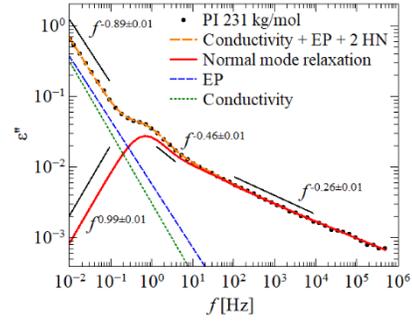

(c) 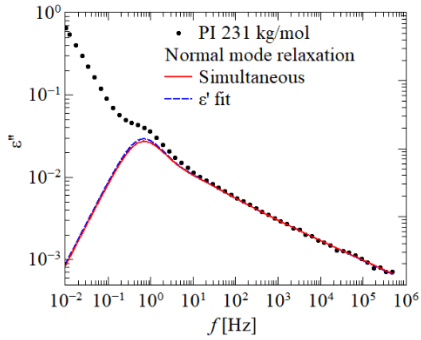

(d) 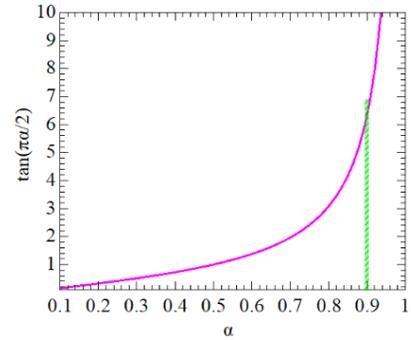

**Figure 5** Dielectric spectra of polyisoprene (PI) with a molecular weight of 231 kg/mol at a temperature 343 K. **(a)** Simultaneous fit of $\epsilon'$ and $\epsilon''$ with conversion factor $\tan\left(\frac{\pi\alpha}{2}\right)$ of EP in $\epsilon''$ **(b)** individual contributions to the normal mode. **(c)** Comparison of $\epsilon''$ from the simultaneous description with $\epsilon''$ calculated from $\epsilon'$. **(d)** Conversion factor as function of the power law exponent $\alpha$. Green shaded area refers to $\alpha = 0.900 \pm 0.002$.

**Table 5** Parameters obtained by fitting the experimental data in Figure 5 with the Sum of 2 HN Functions and a polarization contribution based on Equations (21) and (22). $A_2$ is $A_p \tan\left(\frac{\pi\alpha}{2}\right)$.

|  | $\epsilon_\infty$ | $A_p$ | $\alpha$ | $A_c$ |
|---|---|---|---|---|
|  | $2.30 \pm 0.01$ | $0.0009 \pm 0.0001$ | $0.900 \pm 0.002$ | $0.003 \pm 0.001$ |
| $1^{st}$ HN | $\Delta\epsilon_1$ | $\tau_{HN1}$ (s) | $\beta_1$ | $\gamma_1$ |
|  | $0.060 \pm 0.002$ | $0.28 \pm 0.01$ | $0.999 \pm 0.007$ | $0.80 \pm 0.04$ |
| $2^{nd}$ HN | $\Delta\epsilon_2$ | $\tau_{HN2}$ (s) | $\beta_2$ | $\gamma_2$ |
|  | $0.026 \pm 0.002$ | $0.024 \pm 0.004$ | $0.931 \pm 0.001$ | $0.252 \pm 0.003$ |

Here we used the representation by HN as a potential answer to the calculation of $\epsilon''$ from $\epsilon'$ and showed that a simultaneous fit of $\epsilon'$ and $\epsilon''$ may be the more reliable approach provided that pre-factors are included in the conversion, instead of introducing additional fit parameters. However, each calculated $\epsilon''$ from $\epsilon'$, or $\epsilon''$ from simultaneous fit agree very well, at least within the experimental accuracy. Before we will take further advantage of $\epsilon''_{Brather}$ introduced in the theoretical section, we turn our attention to the higher molecular weight, with even stronger contributions of polarization and conductivity.

## Polyisoprene 436 kg/mol

As Figure 6 illustrates, both real and imaginary part of the complex dielectric permittivity of PI with a molecular weight of 436 kg/mol appear to be relatively featureless. Either strong contributions of conductivity and polarization or both seem to entirely overwhelm the normal mode peak position and maybe strongly affect shape, hence the power law too. The important question to address here is whether conclusions for relaxation time and shape can be drawn from datasets which contain maybe a residue of a peak and what would be the experimental error of the extracted parameters.

From the previous sections we have seen that HN and the approximation $\epsilon''_{Brather}$ can be used for a simultaneous fit or to convert $\epsilon'$ to $\epsilon''$, respectively. Here we start with HN. In this particular case of $\epsilon'$ of PI with 436 kg/mol, the polarization is very strong compared to the normal mode relaxation. Hence, the correct incorporation of numerical pre-factors may play an important role in the determination of the peak position as well.

*Fitting $\epsilon'$ only*

As Figure 6 illustrates, despite the data is well described by $\epsilon'_{HN}$, the calculated $\epsilon''_{HN}$ does not match experimental data. While we would expect strong deviations in the vicinity of the peak position and at low frequencies, the functional dependence is entirely different. This is surprising, because we only apply a conversion via the HN equation which is a consequence of the Kramers-Kronig relation that should theoretically connect real and imaginary part of the complex dielectric function $\epsilon^*$. More importantly, to advance the understanding on the crossover molecular weight, it is important to find out which result, the experimental $\epsilon''$ or the one calculated from $\epsilon'$, represents the real molecular dynamics.

Let's discuss the potential origin of the differences. Either the model including the Kramers-Kronig relation is not appropriate to describe the data, or conductivity and polarization such strongly contribute to $\epsilon''$ that neither the peak position nor the peak shape associated with the normal mode can be reliably determined. This would also imply that neither the peak position nor the peak shape can be used to draw important conclusions on the contributing molecular processes. However, both calculated data and experimental $\epsilon''$ point to a broad relaxation spectrum, with a slowly decaying tail at higher frequencies ($f^{-0.09}$ or $f^{-0.27}$). This observation points to a second relaxation mechanism between reptation and $\alpha$-relaxation and would be compatible with a potential CLF contribution. For the sake of completeness, the relaxation times associated with the normal mode are reported (Table 6) but may be questionable as only a barely visible hump at around 0.1 Hz to 1 Hz seems to indicate a residue of a relaxation peak. We tend not to interpret this hump as the maximum of a relaxation process as it might be eventually misleading.

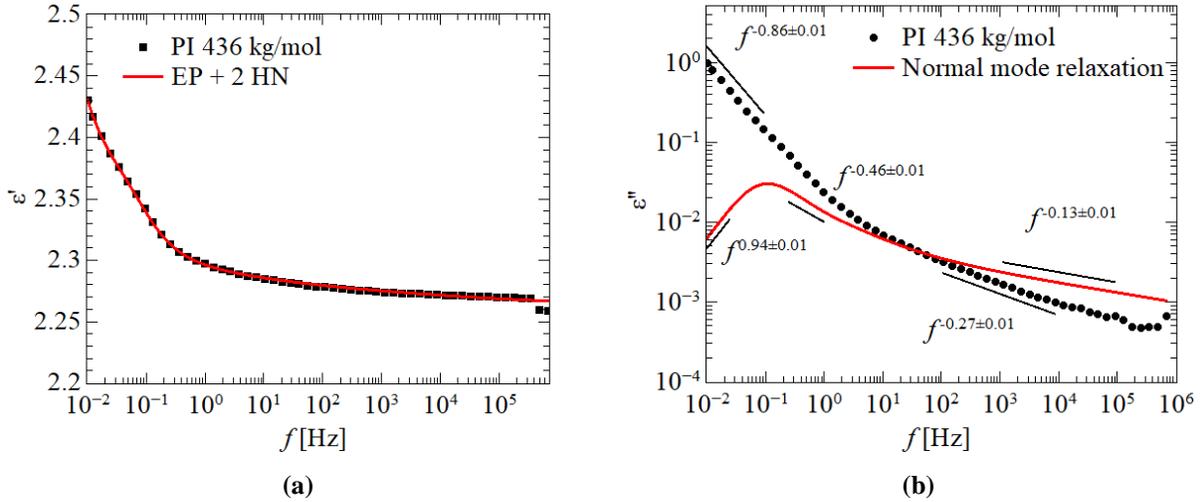

**Figure 6** Dielectric spectra of polyisoprene (PI) with a molecular weight of 436 kg/mol at a temperature 353 K. **(a)** $\epsilon'$ fit by EP and 2 HN functions **(b)** Normal mode relaxation based on the $\epsilon'$ fit parameters

**Table 6** Parameters obtained by fitting the experimental data $\epsilon'$ of polyisoprene with a molecular weight of 436 kg/mol at a temperature 353 K in Figure 6 with the sum of 2 HN functions and an electrode polarization contribution based on Equation (21).

|  | $\epsilon_\infty$ | $A_p$ | $\alpha$ |  |
|---|---|---|---|---|
|  | $2.263 \pm 0.005$ | $0.0005 \pm 0.0002$ | $1.04 \pm 0.06$ |  |
| $1^{st}$ HN | $\Delta\epsilon_1$ | $\tau_{HN1}$ (s) | $\beta_1$ | $\gamma_1$ |
|  | $0.074 \pm 0.004$ | $2.1 \pm 0.2$ | $1.00 \pm 0.02$ | $0.66 \pm 0.05$ |
| $2^{nd}$ HN | $\Delta\epsilon_2$ | $\tau_{HN2}$ (s) | $\beta_2$ | $\gamma_2$ |
|  | $0.027 \pm 0.004$ | $0.13 \pm 0.03$ | $1.00 \pm 0.08$ | $0.12 \pm 0.04$ |

*Simultaneous Fit of $\epsilon'$ and $\epsilon''$*

As we realized in case of the PI with 231 kg/mol the joint fit may lead to more reliable results, we also try this approach here. Figure 7 presents the simultaneous fit results of $\epsilon'$ and $\epsilon''$ spectra and Table 7 summarizes the parameters. Compared to the fit results illustrated in Figure 6, the model function seems to represent the experimental data more accurately, as it describes $\epsilon'$ and $\epsilon''$ equally well.

Furthermore, we show the different contributions using the model functions and parameters from the fit. It helps to better explore the strength of the different contributions compared to the one of $\epsilon''$ (Figure 7 (b)). We distinguish dynamics (in this case the normal mode), polarization, and conductivity. We see two intersections at roughly 0.1 Hz and slightly above 1 Hz, which could already provide an explanation of the barely visible hump in the data without being related to a relaxation time. More importantly, singling out the different contributions indicates that polarization is almost one order of magnitude stronger than both conductivity and $\epsilon''$ at the peak position. This introduces an error for the frequency at the peak maximum or corresponding relaxation time, which needs to be discussed more in detail hereafter. The interpolation of the data at high frequencies comes from the sum of two independent HN functions, hence represent experimental data. This calculation of $\epsilon''$ using the sum of 2 empirical HN functions as a simple interpolation tool reveals power laws $f^{-0.48\pm0.01}$ and $f^{-0.26\pm0.01}$ that reminds very strongly of the reptation ($f^{-0.5}$) and CLF ($f^{-0.25}$) interpretation, respectively.[6]

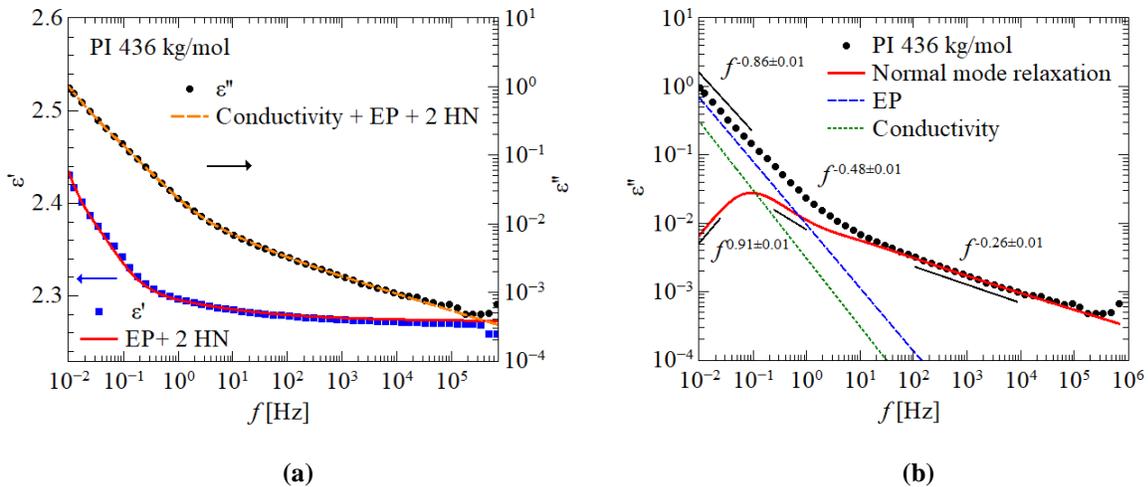

(a)        (b)

**Figure 7** Dielectric spectra of polyisoprene (PI) with a molecular weight of 436 kg/mol at a temperature 353 K. (a) Simultaneous fit of $\epsilon'$ and $\epsilon''$ with conversion factor (b) Normal relaxation, EP contribution and conductivity contribution comparison.

Table 7 Parameters obtained by fitting the experimental data of polyisoprene (PI) with a molecular weight of 436 kg/mol at a temperature 353 K in Figure 7 with the sum of 2 HN functions and polarization contribution based on

|  | $\epsilon_\infty$ | $A_p$ | $\alpha$ | $A_c$ |
| --- | --- | --- | --- | --- |
|  | $2.27 \pm 0.01$ | $0.0011 \pm 0.0002$ | $0.93 \pm 0.01$ | $0.003 \pm 0.001$ |
| $1^{st}$ HN | $\Delta\epsilon_1$ | $\tau_{HN1}$ (s) | $\beta_1$ | $\gamma_1$ |
|  | $0.067 \pm 0.004$ | $2.4 \pm 0.2$ | $0.99 \pm 0.01$ | $0.67 \pm 0.03$ |
| $2^{nd}$ HN | $\Delta\epsilon_2$ | $\tau_{HN2}$ (s) | $\beta_2$ | $\gamma_2$ |
|  | $0.018 \pm 0.001$ | $0.06 \pm 0.01$ | $0.78 \pm 0.03$ | $0.31 \pm 0.01$ |

At this point it is interesting whether we can gain independent information using $\epsilon''_{Brather}$ as a convenient approximate representation of the Kramers-Kronig relation. As this function represents accurate mathematical operations it does not require or introduce pre-factors or does not assume that the spectra are the sum of several underlying contributions. Hence, it is entirely model independent. Furthermore, we can calculate the right-hand side of Brather's equation from $\epsilon'$, which excludes the conductivity. Hence, we should be able to obtain at least some basic information on the strength of conductivity vs the strength of polarization, which then can be compared with the results of the HN fit, displayed in Figure 7 (b).

Before we consider PI with 436 kg/mol, we carefully test the influence of the number of order terms. As Figure 8 (a) illustrates, in case of the low molecular weight PI (12 kg/mol), $N = 20$ terms seem to be sufficient to achieve a reasonable accuracy. Here, the symbols represent calculated values from the previous fit, which allows to separately test the minimum number of modes required for normal mode and polarization more easily. A closer look shows that after convergence is achieved curves do not change, as calculated values are indistinguishable for $N > 20$. Figure 8 (b) shows that $N = 20$ can also be used if the spectrum is broader, such as in higher molecular weight polymers, like PI 436 kg/mol. However, once polarization contributes $N = 100$ terms are required to accurately reflect experimental $\epsilon''$, Figure 8 (c). Deviations are indicated by the $\epsilon''_{experimental} - \epsilon''_{Brather}$ and appear to be almost one order of magnitude smaller and show a power law $f^{-1.00 \pm 0.01}$. The frequency dependence points to conductivity as potential explanation of this residue. Indeed, a comparison of Figures 7 (b) and 8 (b) points to the conductivity as the most likely source of this residue.

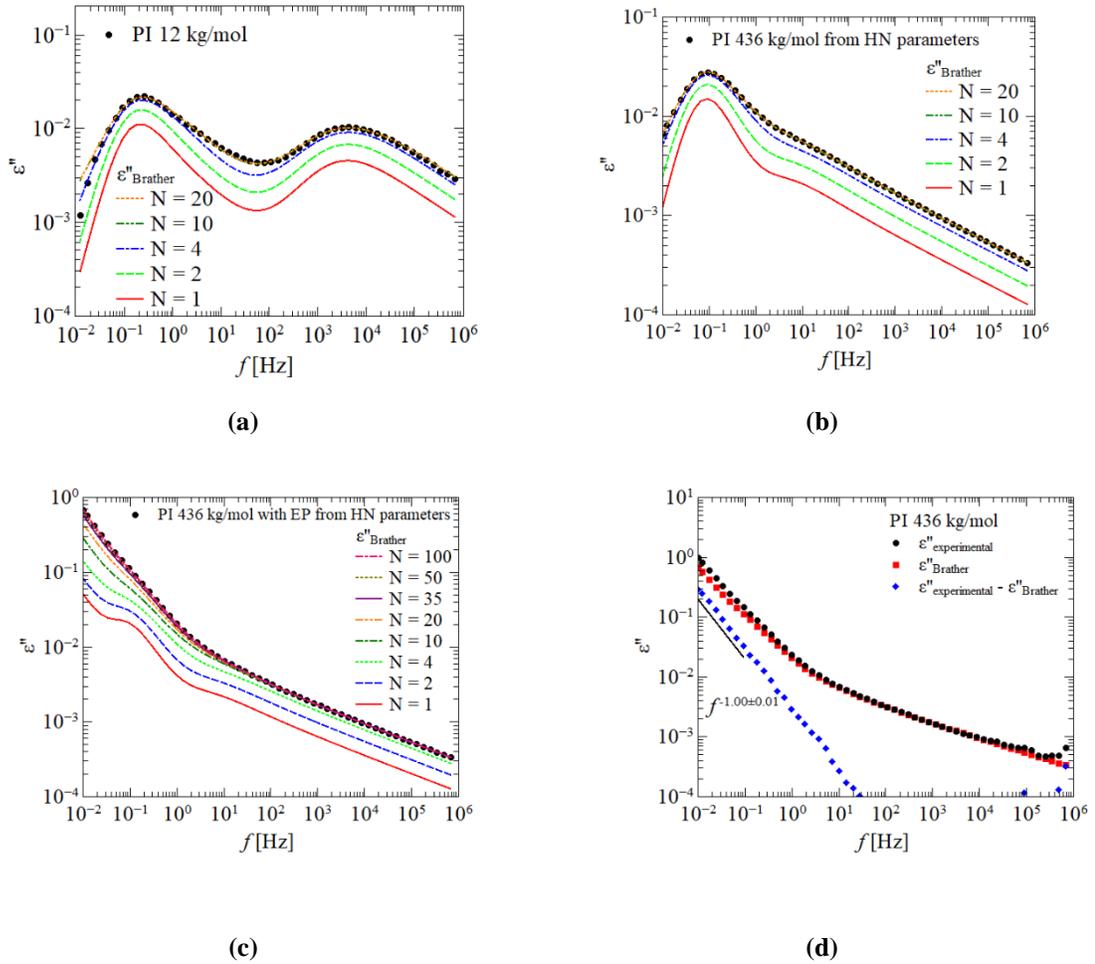

**Figure 8** Illustration of the effect of higher order terms in $\epsilon''_{Brather}$ using **(a)** PI with 12 kg/mol. The terms included in $\epsilon''_{sum}$ goes from $n = 1$ to 20. **(b)** Calculated values for PI with 436 kg/mol, and **(c)** additional contributions by polarization but no conductivity, which has been calculated using the $\tan\left(\frac{\pi\alpha}{2}\right)$ conversion factor and the exponent from the simultaneous fit. Here, convergence requires $N = 100$. **(d)** Experimental $\epsilon''_{experimental}$ compared with $\epsilon''_{Brather}$ ($N = 100$) and residue $\epsilon''_{experimental} - \epsilon''_{Brather}$.

# Conclusion

As relaxation time increases with molecular weight and conductivity and polarization contributions affect the data more strongly, the accuracy of the determined relaxation time goes down. The uncertainty of the peak position can be easily almost half order of magnitude. We note that polarization and conductivity strongly influence both peak position and shape. In an attempt to avoid unwanted contributions, at least from conductivity, often the derivative technique is used to convert $\epsilon'$ to $\epsilon''$, Figure 9 (a), then to determine the peak position. For this comparison polarization and conductivity contributions are omitted. While calculating the derivative means a convenient tool leads to satisfactorily accurate results of regarding the relaxation time, the frequency dependent shape is not very well reflected. As this peak shape contains information on the underlying relaxation mechanism this is a considerable loss of information.

Although the spectral shape is incorrect, Figure 9 (a) suggests that the frequency dependence reflects different underlying processes. Such a result could point to the underlying relaxation mechanisms as well. In fact, data in Figure 9 (a) seems to strongly suggest the existence of at least two processes contributing to the normal mode. This would fit very well to the interpretation in the literature.[6] However, the situation is more complicated. If the spectrum is a superposition of the linear combination of different processes, the derivative method may entirely fail to reflect the underlying processes. Let's assume the conversion of $\epsilon'$ to $\epsilon''$ of a spectrum that contains both the normal mode spectrum and polarization. The respective equation would be a linear superposition that includes pre-factors that need to be converted as well. However, polarization and mode spectrum have a different conversion factor when it comes to the pre-factors. Hence, the simple derivative would yield a spectrum that may not yield correct information on the relaxation mechanism. Figure 9 (a) already includes the correct conversion factor that satisfies the Kramers-Kronig relation. Figure 9 (b) uses the correction calculation of the $\epsilon''$ spectrum from $\epsilon'$. While the spectra look clearly different from the ones in Figure 9 (a), we observe similar features. The simultaneous fit and the fit of $\epsilon'$ results in a difference in the peak position and the peak shape. One of the most remarkable observations is that an apparently well modelled $\epsilon'$ with only slight differences between different approaches can lead to a very different $\epsilon''$. This is illustrated in Figure 9 (a).

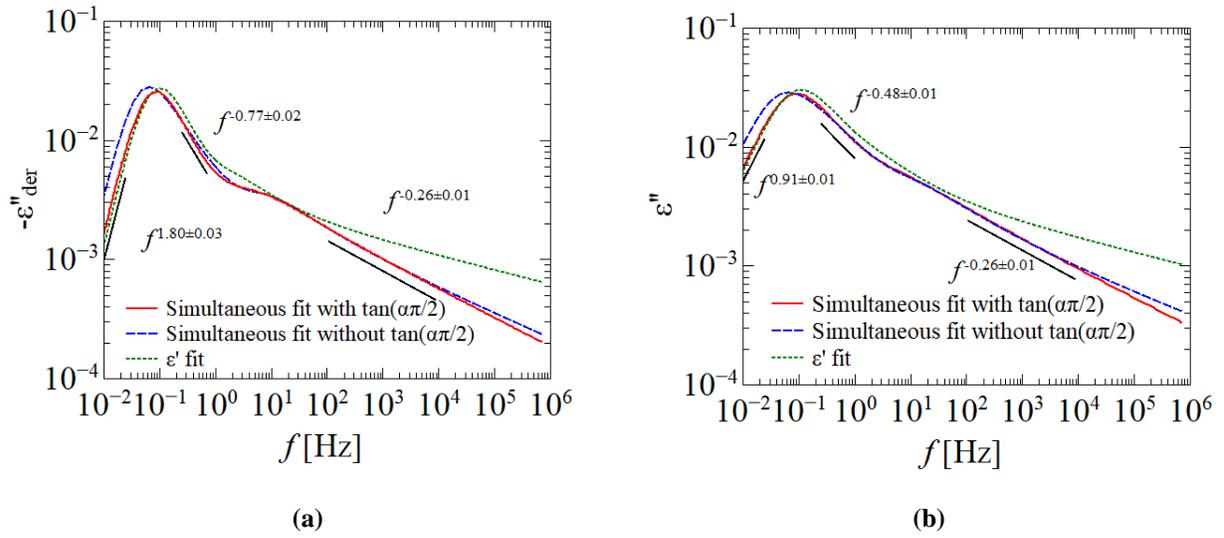

**(a)**                  **(b)**

**Figure 9 (a)** Figure 9 (a) shows the $-\epsilon''_{der}$ of the normal mode relaxation based on the fit parameters obtained from different fitting approaches. Figure **(b)** shows the normal mode relaxation also based on the fit parameters. We see that just taking the derivative would reflect slopes that is different from theoretical and experimental observation (except the high frequency wing which has a good linear behavior and thus the logarithmic derivative does not change the slope)

Knowing that the error with the peak position can be substantial Figure 10 tests whether the molecular weight dependence would be compatible with the existence of reptation and CLF at higher molecular weights. Based on the above analysis, the error involved does not permit to draw a conclusion based on the molecular weight dependence.

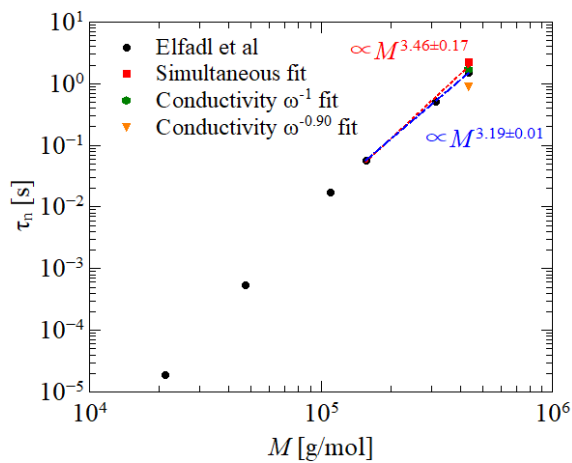

**Figure 10** Molar mass dependence of relaxation time $\tau_n$ at $T = 330$ K, digitized from Elfadl et al.[3] We added the uncertainties associated with the determination shown here. Within the experimental accuracy the data can be described by different power laws as indicated in the figure.

Despite the apparent differences, it also becomes clear that the spectra obtained from the different approaches all indicate a broad relaxation, with a peak from $10^{-2}$ to $10^6$ Hz. Furthermore, the spectra display at least two different processes. If we assume that the peak can be associated with the normal mode, then the reptation model would be compatible with the observed power laws. The long tail for $f \geq 10$ Hz points to contour length fluctuations. While it cannot be fully ruled out that the change of the characteristic frequency dependence of $\epsilon''$ can be caused by polarization and/or conductivity all experimental observations point to a at least 2 different relaxation mechanism visible in the frequency window. This interpretation could also hint to CLF and CR still being active up to the highest molecular weights, e.g., in the example of Matsumiya et al. who studied PI with a molecular weight of 1120 kg/mol.[8,9] While this experimental result still leaves the uncertainty of the transition molecular weight, it can be clearly stated that it is significantly above the one determined from the molecular weight dependence $116 \pm 33$ kg/mol (SI). The experimental data of Matsumiya et al. suggests that the transition molecular weight could even be one order of magnitude higher.

# Acknowledgments

B. J., K.J.B., and G.J.S. gratefully acknowledge funding by the U.S. Department of Energy (DoE) under grant DE-SC0019050. M.W. and G.J.S gratefully acknowledge funding by the U.S. National Science Foundation (NSF) under Award Number 1808059.